\documentclass[aps,showpacs,amsmath,amssymb,reprint,twocolumn,graphicx,superscriptaddress]{revtex4}
\usepackage{graphicx}
\usepackage{dcolumn}
\usepackage{latexsym,bm}
\usepackage{multirow}

\begin{document}
\title[Version zpf0.1]{Topo-electronic transitions in Sb(111) nanofilm: the interplay between quantum confinement and surface effect}

\author{PengFei Zhang}
\affiliation{State Key Laboratory of Low-Dimensional Quantum Physics and Department of Physics, Tsinghua University, Beijing, 100084, People's Republic of China}

\author{Zheng Liu}
\affiliation{Department of Materials Science and Engineering, University of Utah, Salt Lake City, 84112, USA}

\author{Wenhui Duan}
\affiliation{State Key Laboratory of Low-Dimensional Quantum Physics
and Department of Physics, Tsinghua University, Beijing, 100084,
People's Republic of China}

\author{Feng Liu}
\email{fliu@eng.utah.edu}
\affiliation{Department of Materials Science and Engineering, University of Utah, Salt Lake City, 84112, USA}

\author{Jian Wu}
\email{wu@phys.tsinghua.edu.cn}
\affiliation{State Key Laboratory of
Low-Dimensional Quantum Physics and Department of Physics, Tsinghua
University, Beijing, 100084, People's Republic of China}

\date{\today}
\begin{abstract}
When the dimension of a solid structure is reduced, there will be two emerging effects, quantum confinement and surface effect, which dominate at nanoscale. Based on first-principles calculations, we demonstrate that due to an intriguing interplay between these two dominating effects, the topological and electronic (topo-electronic) properties of Sb (111) nanofilms undergo a series of transitions as a function of the reducing film thickness: transforming from a topological semimetal to a topological insulator at 7.8 nm (22 bilayer), then to a quantum spin hall (QSH) phase at 2.7 nm (8 bilayer), and finally to a normal (topological trivial) semiconductor at 1.0 nm (3 bilayer). Our theoretical findings for the first time identify the existence of the QSH in the Sb (111) nanofilms within a narrow range of thickness and suggest that the Sb (111) nanofilms provide an ideal test bed for experimental study of  topo-electronic phase transitions.

\end{abstract}


\maketitle

Thanks to the development of modern epitaxial-growth technique, the state-of-the-art electronic devices are often made of thin films with good control of film crystallinity and thickness with single atomic layer precision. When the film thickness approaches nanoscale, two emerging effects become important. On one hand, the confinement in the z-direction normal to the film surface significantly alters the behavior of the itinerant electrons in the film by quantizing the electron wave, which manifests not only in film structural \cite{Schulte1976427} and mechanical stability \cite{PhysRevLett.102.166404}, but also in electronic phases, such as magnetism \cite{PhysRevB.42.976} and superconductivity \cite{Science.306.1915}. On the other hand, the termination of a material with a surface leads to a change of the electronic band structure differing from the bulk, with surface states formed only at the atomic layers closest to the surface \cite{Davison1992}. Due to the large surface-volume ratio of a thin film, the surface effects may dominate the overall thin film electronic properties.

Recently, a unique type of surface states is identified in materials with strong spin-orbit coupling (SOC), which is expected to serve as the dissipationless conducting channels in spintronic devices \cite{RevModPhys.82.3045, RevModPhys.83.1057}. Taking the form of a spin-resolved Dirac cone lying within the bulk band gap, the so-called helical surface states are theoretically attributed to the nontrivial $Z_2$ topology of the bulk valence bands, which has in turn triggered a vast number of studies on these topologically nontrivial materials, i.e. the topological insulator (TI) in three dimensions (3D) and the QSH systems in 2D \cite{Bern15122006,Konig02112007,PhysRevLett.98.106803,Nat.PhysZhangHaijun}.

An interesting surface coupling effect is observed in TI nanofilms \cite{Nat.PhysXueQi-Kun,ADMA201000368}. In such films, the helical surface states located on the top and bottom surfaces are sufficiently close to each other in space, so that the coupling between the two surfaces becomes noticeable. The surface coupling effect opens a gap at the Dirac point on both surfaces and may even lead to topological transitions of the film \cite{PhysRevB.81.041307,PhysRevB.81.115407}.

One significant difference between the quantum confinement and surface coupling is that they exhibit different scaling laws as a function of the film thickness. Following the textbook description for quantum particles in a box, we have:

\begin{equation}
{\Delta}E_{B}{\sim}\frac{1}{m^{*}L^{2}},
\end{equation}
where ${\Delta}E_{B}$ is the energy shift of a bulk state due to the quantum confinement, $m^{*}$ and $L$ stand for the effective mass of charge carrier and the film thickness, respectively.

The wavefunction of surface states takes the form of $\Psi_{SS}=u(\textbf{r})e^{ik_{x}x}e^{ik_{y}y}e^{-{\lambda}z}$, which modifies the general Bloch wavefunction in a periodic solid by replacing $k_z$ with a complex number $\lambda$. The real and imaginary part of $\lambda$  are the decay constant and oscillating wave vector, respectively. The surface-coupling-induced splitting ${\Delta}E_{S}$ is determined by $\langle\Psi_{SS}^{t}\mid\Psi_{SS}^{b}\rangle$, where $t$ and $b$ stand for the top and bottom surfaces, respectively. Assuming the two surfaces are equivalent and by smoothing out u(\textbf{r}) as a constant, we have

\begin{equation}
{\Delta}E_{S}{\sim}e^{-Re(\lambda) L}
\end{equation}

When the thickness of a thin film is decreased, the quantum confinement with the ``flatter'' power-law scaling is expected to show up first, modifying the bulk bands. The surface coupling will dominate in the 2D limit, because of the ``steeper'' exponential scaling. A joint action of the quantum confinement and surface coupling effect is likely to spawn rich physics in a nanofilm, especially when nontrivial topology is also involved.

In this Letter, using first-principles calculations, we report our finding of a series of topo-electronic transitions in Sb (111) nanofilms triggered by the interplay of these two effects. As the film thickness reduces, the quantum confinement first opens a gap in the bulk band to induce a transition from a topological semimetal to a topological insulator. Then, the surface coupling effect kicks in, opening a gap in the surface bands giving rise to a QSH phase within a narrow range of thickness. The film finally degrades into an ordinary semiconductor in the 2D limit, because of the surface-coupling induced level crossings.

Antimony is one of the main group elements with strong SOC. It crystallizes in a rhombohedral structure as shown in Fig. 1a, which is typical among the group-V solids. Each atom has three equidistant nearest-neighbour (NN) atoms and three equidistant next-NNs slightly farther away. The atoms form a bilayered (BL) structure along the (111) direction (Fig. 1b). Each bilayer has a puckered honeycomb lattice with two atoms per unit cell, corresponding to the two NN atoms. A Sb (111) film consists of several BLs stacking up in an ABC sequence along the (111) direction.

\begin{figure}[ht]
\includegraphics[width=0.5\textwidth]{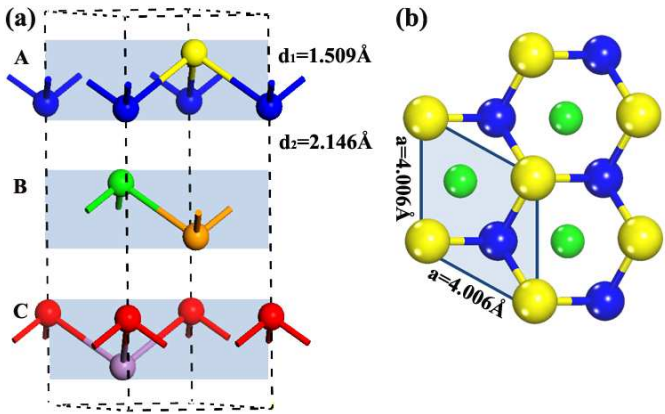}
\caption{\label{Tab 1} (Color online) (1) The hexagonal unit cell of single crystal Sb. (b) The top view of the Sb lattice. The colors and sizes of Sb atoms indicate different atomic layers. }
\end{figure}

Bulk Sb has been predicted to be an intrinsic topological semimetal \cite{PhysRevB.76.045302}, which means that despite a negative indirect band gap, the valence bands of bulk Sb are topologically nontrivial as those of a TI. The helical surface bands in a 20-BL Sb film have been experimentally observed using the angle-resolved photoemission spectroscopy (ARPES) \cite{PhysRevLett.107.036802}. First-principles calculations have also predicted a surface-coupling-induced gap in the 4-BL and 5-BL films \cite{arXiv:1201.1976}. However, a systematic study on all possible topo-electronic transitions in a Sb (111) nanofilm is still missing. Here, we map out its complete topo-electronic phases as a function of thickness, including the discovery of the QSH phase in a narrow thickness regime (1.3-2.7 nm), that is unknown before.

Our calculations are performed with the density functional theory (DFT) using the plane wave basis, as implemented in the ABINIT package \cite{Gonze2005220}. We employ the local density approximation (LDA) \cite{PhysRevB.23.5048} and the Hartwigsen-Goedecker-Hutter pseudopotential \cite{PhysRevB.58.3641}, which is generated on the basis of a fully relativistic all-electron calculation and have been tested to be accurate for heavy elements like Sb. The spin-orbit coupling is incorporated in the self-consistent calculations as described in \cite{Gonze2002478}. The reliability of the standard DFT+LDA calculations on group-V semimetals including bulk Sb has been carefully examined by comparing with experimental data for the density of states, number of free carriers and Fermi surface \cite{PhysRevB.41.11827}. The numerical accuracy has been shown in strict control, with unexpected accuracy even for the Fermi surface, which is sensitive to the energy gap.

Our calculated structural parameters of bulk Sb are shown in Fig. 1a and 1b, in good agreement with previous calculations \cite{PhysRevB.41.11827}. To model the thin film, a supercell of slab is used with periodic boundary conditions in all three dimensions and a 10 \AA ~ vacuum layer between the slabs to eliminate the inter-slab interaction. The thin film has slightly relaxed structural parameters relative to the bulk values. All the atomic positions are fully relaxed as the film thickness changes. A plane wave cut off of 28 Ry and a $\Gamma$-centered k-point mash of 8$\times$8$\times$1 are used.

\begin{figure}[ht]
\includegraphics[width=0.48\textwidth]{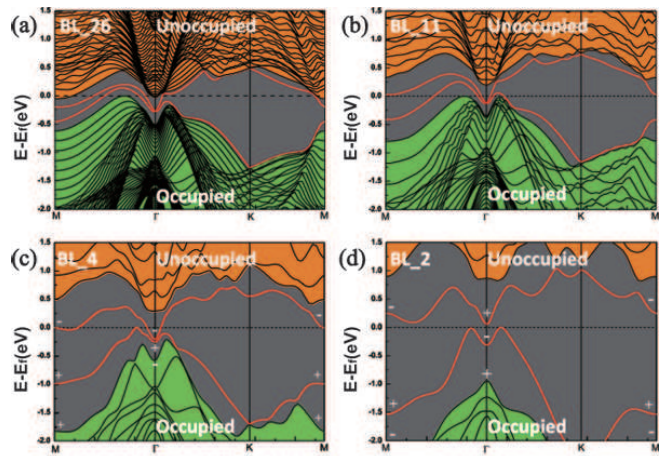}
\caption{\label{Tab 2} (Color online) The band structure of Sb (111) films with different thickness. (a) 26-BL, topological semimetal with a negative bulk gap; (b) 11-BL, TI phase; (c) 4-BL, QSH phase ; (d) 2-BL, ordinary semiconductor. The Fermi level is set to zero. For the 4-BL and 2-BL films, parity information at the $\Gamma$ and M points is shown to distinguish their topological difference. }
\end{figure}

In Fig. 2a, we plot the band structure of a 26-BL (9.4 nm) film, which can be well understood as the bulk band projections (shaded regions) plus the surface bands (red curves) in the middle of the band gap. The conduction band edge and the valence band edge overlap at the Fermi level, showing the typical feature of a semimetal. A pair of surface bands cross at the $\Gamma$-point, forming a Dirac cone. The calculated band structure is in good agreement with recent ARPES data and earlier calculations \cite{PhysRevLett.107.036802,arXiv:1201.1976}, which is consistent with that fact that bulk Sb is a topological semimetal.

We now reveal how the quantum confinement and surface coupling effect modify the band structure when the film thickness decreases. For a semimetal, the dominating effect of quantum confinement is to enlarge the energy gap between the bulk conduction bands and valence bands due to the opposite signs of $m^{*}$ for electrons and holes. Two kinds of bulk gaps are traced in Fig. 3 as a function of the film thickness. One is the direct gap at the $\Gamma$-point; the other is the indirect gap between the conduction band minimum and the valence band maximum. Both gaps are measured with respect to the bulk states excluding the two surface bands.

The increase of both the direct (squares) and indirect gaps (diamonds) roughly follows the $1/L^{2}$ scaling as expected, where L is the thickness. An important transition occurs at 22-BL, where the indirect gap reverses its sign from negative to positive, signifying a semimetal-to-semiconductor transition, while the direct gap keeps open. Since the transition does not involve any level switching between the conduction and valence bands, the topology of the film does not change. Therefore, at 22-BL, the film transforms from a topological semimetal to a TI driven by the quantum confinement of bulk electronic states. A typical band structure of the Sb TI film is plotted in Fig. 2b.

\begin{figure}[h]
\includegraphics[width=0.5\textwidth]{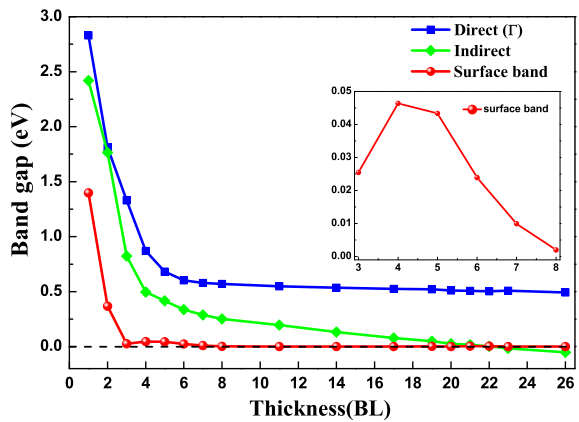}
\caption{\label{Tab 3} (Color online) The direct ($\Gamma$ point) and
indirect bulk gap and surface splitting as a function of the film thickness.
Inset:the enlarged plot of the surface band splitting from 3-BL to
8-BL. }
\end{figure}

In Fig. 3, we have also plotted the splitting (dots) of the Dirac point formed by the two surface bands, which reflects the surface coupling effect. The Dirac point splitting becomes noticeable only at a thickness much smaller than the semimetal-to-semiconductor transition point. This is because the exponential scaling of the surface coupling decays much faster than the power-law decay of quantum confinement. It is interesting to note that the splitting curve is not monotonic: first increases slowly, reaching the maximum at 4-BL; then drops to a minimum at 3-BL; finally increases rapidly from 3-BL to 1-BL (Fig. 3). A magnified view of the turn-around from 8-BL to 3-BL is shown as the inset of Fig. 3. Such an anomaly has also been observed in earlier first-principles study \cite{arXiv:1201.1976} , which attributed this behavior to the inversion-symmetry breaking. In our calculation, however, the inversion symmetry is preserved for all the films. Therefore, some other mechanisms should be considered.

Note that when deriving Eq. (2), we arbitrarily set the periodic Bloch function u(\textbf{r}) to be a constant. If the details in u(\textbf{r}) are retained, Eq. (2) should be rewritten as:

\begin{equation}
{\Delta}E_{S}{\sim}e^{-{\lambda}L}\int_0^L \mathrm{d}z u^{*}(L\widehat{z}-\textbf{r})u(\textbf{r})
\end{equation}
Here, we still assume that the two surfaces are equivalent and thus the in-plane components can be integrated out. Simply because of the rapid oscillating nature of the Bloch function, the additional integral in Eq. (3) has to be essentially non-monotonic as a function of L, varying not only in the magnitude but also in the phase.

Besides a minor turnover as shown in Fig. 4, the magnitude modification does not override the overall exponential trend. The phase modification, however, may result in more significant changes in terms of the band topology \cite{RevModPhys.82.3045, RevModPhys.83.1057}. Specifically, for the Dirac point, u(\textbf{r}) can be chosen to be real, because it locates at the $\Gamma$-point. The phase, which is the sign in this specific case, of ${\Delta}E_{S}$ in principle determines the relative positions of the bonding and anti-bonding levels formed by the coupled top and bottom surface states. It is known that a switching between an occupied level and an unoccupied level with the opposite parity is going to change the overall topology of the occupied bands \cite{PhysRevB.76.045302}.

If we view surface-coupling-induced splitting at the Dirac point as the energy gap of a 2D system, the turnover of the surface splitting shown in Fig. 4 is analogous to a gap closing-reopening process, which is the precursor of topological transitions \cite{RevModPhys.82.3045, RevModPhys.83.1057}. To demonstrate this idea, we have calculated the $Z_2$ topological invariant of Sb (111) film from 1-BL to 7-BL, which serves as the ``order parameter'' to differentiate the topologically trivial and nontrivial 2D phases. The calculation of the $Z_2$ invariant can be dramatically simplified by the so-called  ``parity method'' \cite{PhysRevB.76.045302}, if the system is space inversion invariant, as the case of the Sb (111) film. Accordingly, the $Z_2$ number of the Sb film can be obtained from the wavefunction parities at four time-reversal-invariant k-points, ($K_i$), one $\Gamma$ and three M's, as:

\begin{equation}
\delta(K_{i})=\prod_{m=1}^{N}\xi^{i}_{2m}, \ \ (-1)^{\nu}=\prod_{i=1}^{4}\delta(K_{i})=\delta(\Gamma)\delta^{3}(M)
\end{equation}
where $\xi=\pm1$ is the parity eigenvalues and N is the number of the occupied bands. The $Z_2$ topological invariant $\nu$ takes two values: $\nu=1$ indicating a nontrivial phase and $\nu=0$ indicating a trivial phase.

\begin{table}[ht]
\caption{\label{Tab 4}The total parity at the $\Gamma$ and M points
and the $Z_2$ number of Sb(111) films with different thickness.}
\begin{ruledtabular}
\begin{tabular}{cccccccc}
No.of BLs          &1 &2 &3 &4 &5 &6 &7 \\
\hline
$\delta$($\Gamma$) &$\mathbf{-}$ &$\mathbf{-}$ &+ &$\mathbf{-}$ &$\mathbf{-}$ &+ &+ \\
3$\delta$(M)       &$\mathbf{-}$ &$\mathbf{-}$ &+ &+ &+ &$\mathbf{-}$ &$\mathbf{-}$ \\
$\nu$              &0 &0 &0 &1 &1 &1 &1 \\
\end{tabular}
\end{ruledtabular}
\end{table}

Table \textrm{I} presents the calculated $Z_2$ topological invariants. Clearly, a phase transition occurs between 3-BL and 4-BL, which is in agreement with the speculation above. Between 1-BL and 3-BL, the films are topologically trivial. Between 4-BL and 7-Bl, the films are topologically nontrivial, leading to a QSH phase. The calculated surface band splitting of 4-BL and 5-BL is above 30meV, providing a robust QSH gap at room temperature. Typical band structures of the trivial semiconductor phase and the QSH phase are plotted in Fig. 2c and 2d, respectively.

\begin{figure}[h]
\includegraphics[width=0.45\textwidth]{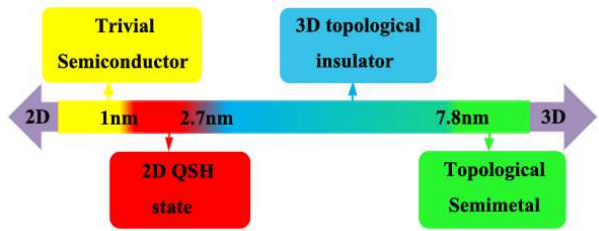}
\caption{\label{Tab 5} (Color online) The phase transition diagrams
as a function of the film thickness from trivial semiconductor to
topological semimetal.}
\end{figure}

We should point out that the topology of the multi-BL Sb films cannot be interpreted as that of a stack of 1-BLs. Otherwise, all the Sb films would be topologically trivial because the 1-BL film is topologically trivial as found in Tab. I. The outcome of the QSH phase is understandable after considering the level crossings induced by the inter-BL coupling as discussed for the Bi(111) films \cite{PhysRevLett.107.136805}.

In conclusion, by a systematic study of the band structure and wavefunction parity as a function of thickness, we now have a complete understanding of the topo-electronic properties of Sb (111) films acrossing from the limit of 2D film to 3D bulk. As summarized in Fig. 4, in the ``parameter space" of thickness, the topo-electronic phases can be divided into four distinct regimes. The existence of such a rich topo-electronic phase diagram, underlied by a delicate interplay between the quantum confinement and the surface effect, makes Sb (111) nanofilms an ideal test bed for experimental investigation of topo-electronic phase transitions.

We would like to thank M.Y. Chou for stimulating our calculations on Sb, and C.X. Liu for valuable discussions. The work is supported by the Ministry of Science and Technology of China (Grants No. 2011CB606405, No. 2011CB921901, and No. 2009CB929401) and NSF of China (Grants No. 10974110 and No. 11074139). F. Liu thanks support from the DOE-BES (DE-FG02-04ER46027, 04ER46148) program.

\providecommand{\noopsort}[1]{}\providecommand{\singleletter}[1]{#1}%

\end{document}